\documentclass[a4paper,12pt]{article}
\pdfoutput=1
\usepackage{amssymb}
\usepackage{tipa}
\usepackage{mathrsfs}
\usepackage{bbm}
\usepackage{epsf,amsmath}
\usepackage[dvipsnames,usenames]{color}
\usepackage{graphicx,subfigure}
\usepackage{booktabs,multirow,makecell}
\usepackage{cite}

\newlength{\dinwidth}
\newlength{\dinmargin}
\def\be{\begin{equation}}
\def\ee{\end{equation}}
\def\bal{\begin{align}}
\def\eal{\end{align}}

\begin{document}
\title{\bf Study of Nonleptonic $B^{\ast}_{(s)}\to M_1 M_2$ $(M=D$, $D_s$, $\pi$, $K)$ Weak Decays with Factorization Approach}
\author{
Qin Chang$^{a,b,c}$, Pan-Pan Li$^{a}$, Xiao-Hui Hu$^{a}$ and Lin Han$^{a}$\\
{ $^a$\small Institute of Particle and Nuclear Physics, Henan Normal University, Henan 453007, P.~R. China}\\
{ $^b$\small Institute of Particle Physics, Central Normal University, Wuhan 430079, P.~R. China}\\
{ $^c$\small State Key Laboratory of Theoretical Physics, Institute of Theoretical Physics,}\\[-0.2cm]
{     \small Chinese Academy of Sciences, P.~R. China}
}
\date{}
\maketitle
\begin{abstract}
Motivated by the experiments of heavy flavor physics at running LHC and upgrading SuperKEKB/Belle-II in the future, the nonleptonic $B^{\ast}_{(s)}\to M_1 M_2$ $(M=D$, $D_s$, $\pi$, $K)$ weak decays are studied in this paper. The amplitudes are calculated with factorization approach, and the transition form factors $A_0^{B^{\ast}_{(s)}\to M_1}(0)$ are evaluated within BSW model. With the reasonable approximation $\Gamma_{tot}(B^*_{(s)})\simeq \Gamma(B^*_{(s)}\to B_{(s)}\gamma)$, our predictions of branching fractions are presented. Numerically, the CKM-favored tree-dominated $\bar{B}^{*0} \to D^+D^-_s$ and $\bar{B}^{*0}_s \to D^+_sD^-_s$ decays have the largest branching fractions of the order $\sim{\cal O}(10^{-8})$, and hence will be firstly observed by forthcoming Belle-II experiment. However, most of the other decay modes have the branching fractions~$<{\cal O}(10^{-9})$ and thus are hardly to be observed soon. Besides, for the possible detectable $B^{\ast}_{(s)}$ decays with branching fractions $\gtrsim{\cal O}(10^{-9})$, some useful ratios, such as $R_D$ {\it et al.,} are presented and discussed in detail.
 \end{abstract}
\noindent{{\bf Keywords:} $B^{\ast}$ meson; weak decay; factorization }\\
\noindent{{\bf PACS numbers:} 13.25.Hw\,,  12.39.St }
\newpage
 \section{Introduction}
The heavy flavor physics, such as beauty  and charm physics, offers an important tool to test the Standard Model~(SM), explore the source of CP violation and search for the indirect hints of new physics at the low energy scale. Thanks to the fruitful running of B factories (BaBar and Belle) and Tevatron~(CDF and D0) in the past years, most of B meson decay modes with branching fractions $\gtrsim {\cal O}(10^{-7})$ are well measured, which provides a fertile ground for theoretical study. Especially, some interesting and important phenomena are observed, such as the measured direct CP violation in B system, the ``$\pi K$ and $\pi\pi$ puzzles", CP violating effects related to $B_s-\bar{B}_s$ mixing and so on.  With the running LHCb~\cite{ref:LHCb2013} and upgrading SuperKEKB/Belle-II~\cite{ref:BelleII2011}, the experimental analysis of the heavy flavor physics will be pushed towards new frontiers of precision. Therefore, besides of B meson, some rare weak decay modes of the other heavy mesons, such as $B^*$~\footnote{For convenience of expression, $B^*$ refers to $B^{*+}(B^{*-})$ and ${B^{*0}}~(\bar{B}^{*0})$ mesons except for special notation. } and $B^*_s$ mesons~($J^P=1^-$~\cite{PDG14})~ {\it et al.}, are expected to be observed in the near future. 

Experimentally, different from $\Upsilon(4S)$ resonance which decays predominantly to simple final states $B^0\bar{B}^0$ and $B^+B^-$, $\Upsilon(5S)$ resonance decays mainly to 10 final states with a pair of $B^{(*)}_{(s)}$ mesons. One of the main goals of the $\Upsilon(5S)$ physics program is to study the decays of $B_s$ meson. While, it should be noted that a plenty of $B^{*}_{(s)}$ samples would be produced simultaneously, which can be seen from the measurements of the branching fractions of $\Upsilon(5S)$ decays summarized in Table~\ref{tab:ups}. Some data of $e^+e^-$ collisions at the $\Upsilon(5S)$ resonance have been accumulated by CLEO and Belle collaborations~\cite{Huang:2006em,Louvot:2008sc}, and the masses of $B^*_{(s)}$ mesons have been well measured~\cite{Louvot:2008sc,Louvot:2009ie,Aquines:2006qg,Aaij:2012uva}. However, the integrated luminosity is not high enough to probe $B^*_{(s)}$ rare decays, and there is no available experimental measurement until now. 

Fortunately, with the target luminosity $8\times10^{35}{\rm cm}^{-2}{\rm s}^{-1}$ at forthcoming superB factory SuperKEKB, the annual integrated luminosity is expected to be about $13\,ab^{-1}$ in 2018~\cite{Abe:2010gxa}. With the cross section of $\Upsilon(5S)$ production in $e^+e^-$ collisions, $\sigma(e^+e^-\to\Upsilon(5S))=(0.301\pm0.002\pm0.039)\,{\rm nb}$~\cite{Huang:2006mf}, it is expected that about $4\times10^9$ $\Upsilon(5S)$ samples could be produced per year. Using the data given in Table~\ref{tab:ups}, one may further evaluate roughly the number of $B^{*}_{(s)}$ mesons could be collected by Belle-II per year, that $N(B^{*}+\bar{B}^{*})/{\rm year}\sim4\times10^9$ and $N(B^{*}_s+\bar{B}^{*}_s)/{\rm year}\sim2\times10^9$, which implies that the rare $B^{*}_{(s)}$ meson decays with branching fractions $\gtrsim{\cal O}(10^{-9})$ are hopeful to be observed in the near future. Moreover, due to the much larger beauty production cross section of $pp$ collisions compared with the one of $e^+e^-$ collisions, LHC~(LHCb) also possibly provides some experimental information of  $B^{*}_{(s)}$ meson rare decays~\cite{Aaij:2010gn,Aaij:2014jba}. 

\begin{table}[t] \small
\caption{The branching fractions of $\Upsilon(5S)$ decays related to $B^*$ final states~\cite{PDG14}.}
\centering  \vspace{0.5cm}\label{tab:ups}
\begin{tabular}{c|cccc|cccccccccc}
\hline\hline
Decay Modes & $B\bar{B}^*+c.c$ & $B^*\bar{B}^*$ & $B^*\bar{B}\pi+B\bar{B}^*\pi$ & $B^*\bar{B}^*\pi$ & $B_s\bar{B}^*_s+c.c.$ & $B^*_s\bar{B}^*_s$    \\
\hline
${\cal B}[\%]$ & $13.7\pm1.6$  & $38.1\pm3.4$  & $7.3\pm2.3$ & $1\pm1.4$ & $1.35\pm0.32$ & $17.6\pm2.7$\\
\hline\hline
\end{tabular}
\end{table}

In the past years, because $B^{*}_{(s)}$ meson decay occur mainly through  radiative processes $B^{*}_{(s)}\to B_{(s)}\gamma$ and  their weak decays are generally much rare,  there are few theoretical studies of $B^{*}_{(s)}$ weak decays before. While, because of the rapid development of experiment as just mentioned, the detailed theoretical studies of  $B^{*}_{(s)}$ rare decays are in fact worthwhile then. Recently, some semileptonic  $B_c^*$ meson decays are evaluated with QCD sum rules in Refs.~\cite{Wang:2012hu,Bashiry:2014qia,Zeynali:2014wya}, in which the branching fractions of the order~$ {\cal O}(10^{-8})$ are predicted. Compared with the $B_c^*$ system, $B_{u,d,s}^*$ meson decays are much easier to be measured by future Belle-II experiment due to the fact that the production fractions $f_{u,d,s}$ are generally much larger than $f_c$~\cite{PDG14}. So, the $B_{u,d,s}^*$ meson rare decays are much worthier of being studied. In this paper, we will pay our attention to the nonleptonic $B^{\ast}_{(s)}\to M_1 M_2$ $(M=D$, $D_{s}$, $\pi$, $K)$ decays.  

 Our paper is organized as the following. In section~2, after a brief review of the effective Hamiltonian and factorization approach, the amplitudes of $B^{\ast}_{(s)}\to M_1 M_2$ decays are calculated. In sections~3, the numerical results and discussions are presented in detail. Finally, we draw conclusion in section~4. 

\section{Theoretical Framework}
%In order to evaluate the nonleptonic heavy meson decays, one uses low-energy effective Hamiltonians, which are gotten by making use of the operator product expansion. 
Within the SM, the effective weak Hamiltonian responsible for $b\to p~(p=d,\,s)$ transitions is given as~\cite{ref:Buras1,ref:Buras2}
%%%%%%%%%%%%%%%%%%%%%%%%%%%%%%%%%%%%%%%%%%%%%%%%%
\begin{align}\label{eq:eff}
 {\cal H}_{\rm eff} &= \frac{G_F}{\sqrt{2}} \sum_{q, q^{\prime} = u, c}\biggl[V_{qb}
 V_{q^{\prime}p}^* \sum_{i = 1}^{2}C_i(\mu) O_i (\mu) +V_{qb} V_{qp}^*\sum_{i = 3}^{10}
 C_i(\mu) O_i (\mu) \biggl] +
 {\rm h.c.},
\end{align}
%%%%%%%%%%%%%%%%%%%%%%%%%%%%%%%%%%%%%%%%%%%%%%%%%
where $V_{qb} V_{q^{(\prime)}p}^{\ast}$~($q^{(\prime)}=u$, $c$) are products of the Cabibbo-Kobayashi-Maskawa~(CKM) matrix elements; $\mu\sim m_b$ is the renormalization scale; $O_i$ are the relevant local four-quark operators, whose explicit forms could be found, for instance, in Ref.~\cite{ref:Buras1,ref:Buras2}; $C_{i}$ are corresponding Wilson coefficients, which describe the short-distance contributions and could be calculated perturbatively~(see Ref.~\cite{ref:Buras1,ref:Buras2} for detail for this part). 

With the effective Hamiltonian given above, the amplitude of $B^*\to M_1M_2$ decay~(for the case of $B^*_s$ decay, one replaces $B^*$ by $B^*_s$ in this section) could be expressed as $\langle M_1M_2 |H_{eff}|B^{\ast}\rangle$. To deal with the hadronic matrix element $\langle M_1M_2 |O_i|B^{\ast}\rangle$  involved in amplitude, Naive factorization~(NF) approach~\cite{Fakirov:1977ta,Bauer:1984zv,Wirbel:1985ji,Bauer:1986bm} based on the color transparency mechanism~\cite{Bjorken:1988kk,Jain:1995dd} is explored, and widely used to evaluate meson decays. In the factorization framework, the hadronic matrix element could be factorized as 
\begin{align}
\langle M_1M_2|Q_i|B^*\rangle\simeq\langle M_2|J_2|0\rangle\langle M_1|J_1|B^*\rangle\,,
\end{align}
in which, the final-state meson that carries away the spectator quark from $B^*$ meson is called as $M_1$, and the other one is called as $M_2$. The two current matrix elements $\langle M_2|J_2|0\rangle$ and $\langle M_1|J_1|B^*\rangle$ can be further parameterized  by decay constants and transition form factors. For the case of pseudo-scalar final states, with the definition and convention in Ref.~\cite{Beneke:2000wa}, they read 
\begin{align}
 \langle M_2(p_2)|\bar{q}_1\gamma^{\mu}\gamma_5q_2|0\rangle=&-if_{M_2}p_2^{\mu}\,,\\
\langle M_2(p_2)|\bar{q}_1i\gamma_5q_2|0\rangle=&f_{M_2}\mu_{M_2}\,,\\
 \langle M_1(p_1)|\bar{q}_3\gamma^{\mu}b|B^{\ast}(p)\rangle=&-\frac{2iV(q^2)}{m_{B^{\ast}}+m_{M_1}}\epsilon^{\mu\nu\rho\sigma}\varepsilon_{\nu}p_{\rho}p_{1{\sigma} }\,, \\
\langle M_1(p_1)|\bar{q}_3\gamma^{\mu}\gamma_5b|B^{\ast}(p)\rangle= &2m_{B^{\ast}}A_0(q^2)\frac{\varepsilon\cdot q}{q^2}q^{\mu}+(m_{M_1}+m_{B^{\ast}})A_1(q^2)(\varepsilon^{\mu}-\frac{\varepsilon\cdot{q}}{q^2}q^{\mu}) \nonumber \\
&+A_2(q^2)\frac{\varepsilon\cdot q}{m_{M_1}+m_{B^{\ast}}}[(p+p_1)^{\mu}-\frac{(m_{B^{\ast}}^2-m_{M_1}^2)}{q^2}q^{\mu}]\,,
 \end{align}
where $q=p-p_1$, $\mu_{M_2}=m_{M_2}^2/(m_{q_1}+m_{q_2})$ and the sign convention $\epsilon^{0123}=-1$.

 Even though some improved approaches, such as the QCD factorization (QCDF)~\cite{Beneke1,Beneke2}, the perturbative QCD (pQCD)~\cite{KLS1,KLS2} and the soft-collinear effective theory (SCET) \cite{scet1,scet2,scet3,scet4}, are presented to evaluate higher order corrections of QCD and reduce the renormalization scale dependence, the NF approach is also a useful tool as a rough theoretical estimation. Because there is no available experimental measurement until now and the decay modes considered in this paper are tree-dominated, the NF approach is enough as a primary analysis and adopted in our evaluation. In addition, within the QCDF at the lowest order, the NF is recovered ~\cite{Beneke1,Beneke2}.

Before presenting the amplitudes of non-leptonic two-body $B^*$ decays, for convenience of expression, we would like to define some quantities with generic form, which are similar to the ones of $B$ meson decays given in Ref.~\cite{Beneke:2003zv}. The effective coefficients $\alpha_i$ of the flavor operators is defined as follows:
\begin{align}\label{eq:a12}
\alpha_{1\,,2}(M_1M_2)&=a_{1\,,2}(M_1M_2)\,,\\
\label{eq:a3}
\alpha_3(M_1M_2)&=a_3(M_1M_2)-a_5(M_1M_2)\,,\\
\label{eq:a3ew}
\alpha_{3,EW}(M_1M_2)&=a_9(M_1M_2)-a_7(M_1M_2)\,,\\
\label{eq:a4}
\alpha_4(M_1M_2)&=a_4(M_1M_2)-r^{M_2}_\chi a_6(M_1M_2)\,,\\
\label{eq:a4ew}
\alpha_{4,EW}(M_1M_2)&=a_{10}(M_1M_2)-r^{M_2}_\chi a_8(M_1M_2)\,.
\end{align}
The general form of the coefficient $a_i$ within the NF framework is defined as 
\begin{align}
a_i(M_1M_2)=C_i+\frac{C_{i\pm1}}{{N_c}}\,,
\end{align}
in which the upper~(lower) sign applies when $i$ is odd~(even). The ratio $r^{M_2}_\chi$ is the so-called chirally-enhanced factor, which is defined as 
\begin{align}
r^{M_2}_{\chi}=\frac{2m_{M_2}^2}{(m_b(\mu)+m_{q_3}(\mu))(m_{q_1}(\mu)+m_{q_2}(\mu))}\,,
\end{align}
where $q_{1,2}$ are the (anti-)quark components in $M_2$, and $q_3$ is the non-spectator quark contained in $M_1$.
 In addition, the generic form of quantity $A_{M_1M_2}$ is given by
\begin{align}
A_{M_1M_2}=i\frac{G_F}{\sqrt{2}}(-2m_{B^{\ast}})(\varepsilon \cdot p_{M_2})A_0^{B^{\ast}\to M_1}(0)f_{M_2}\,,
\end{align}
where $\varepsilon$ is the polarization four-vector of the $B^{\ast}$ meson, $A_0^{B^{\ast}\to M_1}(0)$ the form factor, and $f_{M_2}$ the decay constant. 

With the above definitions, we then present the amplitudes of two-body nonleptonic $B^{\ast}_{u,d,s}\to D_{(s)}D_{(s)}\,, \pi D_{(s)}$ and $K D_{(s)}$ decays:
\begin{itemize}
\item $B^{\ast}_{u,d}\to D_{(s)}D_{(s)}$ decays:
\begin{align}
\label{eq:bmd0dm}
{\cal A}(B^{*-} \to D^0D^-)&=A_{D^0D^-}\big[V_{cb}V_{cd}^*\alpha_1+\sum_{q=u,c}V_{qb}V_{qd}^*(\alpha_4+\alpha_{4,EW})\big]\\
\label{eq:b0dpdm}
{\cal A}(\bar{B}^{*0}\to D^+D^-)&=A_{D^+D^-}[V_{cb}V_{cd}^*\alpha_1+\sum_{q=u,c}V_{qb}V_{qd}^*(\alpha_4+\alpha_{4,EW})\big]\\
\label{eq:bmd0dsm}
{\cal A}(B^{*-} \to D^0D_s^-)&=A_{D^0D_s^-}\big[V_{cb}V_{cs}^*\alpha_1+\sum_{q=u,c}V_{qb}V_{qs}^*(\alpha_4+\alpha_{4,EW})\big]\\
\label{eq:b0dpdsm}
{\cal A}(\bar{B}^{*0} \to D^+D_s^-)&=A_{D^+D_s^-}\big[V_{cb}V_{cs}^*\alpha_1+\sum_{q=u,c}V_{qb}V_{qs}^*(\alpha_4+\alpha_{4,EW})\big]
\end{align}
In addition, $\bar{B}^{*0}\to D^0\bar{D}^0$ and $D_s^+D_s^-$ decays occurred through annihilation process to QCD next-to-leading order and $\bar{B}^{*0}\to D_s^+D^-$ decay induced by $\Delta d=2$ transition are power suppressed, and hardly to be observed in the near future. So, such decays are not evaluated, neither of the other similar decays in the following.

\item $\bar{B}_s^{*0}\to D_{(s)}D_{(s)}$ decays:
\begin{align}
{\cal A}(\bar{B}_s^{*0}\to D_s^+D^-)&=A_{D_s^+D^-}[V_{cb}V_{cd}^*\alpha_1+\sum_{q=u,c}V_{qb}V_{qd}^*(\alpha_4+\alpha_{4,EW})\big]\\
{\cal A}(\bar{B}_s^{*0}\to D_s^+D_s^-)&=A_{D_s^+D_s^-}[V_{cb}V_{cs}^*\alpha_1+\sum_{q=u,c}V_{qb}V_{qs}^*(\alpha_4+\alpha_{4,EW})\big]
\end{align}

\item $B^{\ast}_{u,d}\to \pi D_{(s)}$ decays:
\begin{align}
\label{BmpimD0}
{\cal A}(B^{*-} \to \pi^-D^0)&=A_{D^0\pi^-}V_{cb}V_{ud}^*\alpha_1+A_{\pi^-D^0}V_{cb}V_{ud}^*\alpha_2\\
{\cal A}(B^{*-} \to \pi^-\bar{D}^0)&=A_{\pi^-\bar{D}^0}V_{ub}V_{cd}^*\alpha_2\\
\sqrt{2}{\cal A}(B^{*-} \to \pi^0D^-)&=A_{\pi^0D^-}V_{ub}V_{cd}^*\alpha_1\\
{\cal A}(\bar{B}^{*0} \to \pi^-D^+)&=A_{D^+\pi^-}V_{cb}V_{ud}^*\alpha_1\\
{\cal A}(\bar{B}^{*0} \to \pi^+D^-)&=A_{\pi^+D^-}V_{ub}V_{cd}^*\alpha_1\\
-\sqrt{2}{\cal A}(\bar{B}^{*0} \to \pi^0D^0)&=A_{\pi^0D^0}V_{cb}V_{ud}^*\alpha_2\\
-\sqrt{2}{\cal A}(\bar{B}^{*0} \to \pi^0\bar{D}^0)&=A_{\pi^0\bar{D}^0}V_{ub}V_{cd}^*\alpha_2\\
\sqrt{2}{\cal A}(B^{*-} \to \pi^0D_s^-)&=A_{\pi^0D_s^-}V_{ub}V_{cs}^*\alpha_1\\
{\cal A}(\bar{B}^{*0} \to \pi^+D_s^-)&=A_{\pi^+D_s^-}V_{ub}V_{cs}^*\alpha_1
\end{align}

\item $\bar{B}_s^{*0}\to \pi D_{(s)}$ decays:
\begin{align}
{\cal A}(\bar{B}_s^{*0} \to \pi^-D_s^+)&=A_{D_s^+\pi^-}V_{cb}V_{ud}^*\alpha_1
\end{align}

\item $B^{\ast}_{u,d}\to K D_{(s)}$ decays:
\begin{align}
{\cal A}(B^{*-} \to K^-D^0)&=A_{D^0K^-}V_{cb}V_{us}^*\alpha_1+A_{K^-D^0}V_{cb}V_{us}^*\alpha_2\\
{\cal A}(B^{*-} \to K^-\bar{D}^0)&=A_{K^-\bar{D}^0}V_{ub}V_{cs}^*\alpha_2\\
{\cal A}(\bar{B}^{*0} \to K^-D^+)&=A_{D^+K^-}V_{cb}V_{us}^*\alpha_1\\
{\cal A}(\bar{B}^{*0} \to \bar{K}^0D^0)&=A_{\bar{K}^0D^0}V_{cb}V_{us}^*\alpha_2\\
{\cal A}(\bar{B}^{*0} \to \bar{K}^0\bar{D}^0)&=A_{\bar{K}^0\bar{D}^0}V_{ub}V_{cs}^*\alpha_2
\end{align}

\item $\bar{B}_s^{*0}\to K D_{(s)}$ decays:
\begin{align}
{\cal A}(\bar{B}_s^{*0} \to K^+D^-)&=A_{K^+D^-}V_{ub}V_{cd}^*\alpha_1\\
{\cal A}(\bar{B}_s^{*0} \to K^0D^0)&=A_{K^0D^0}V_{cb}V_{ud}^*\alpha_2\\
{\cal A}(\bar{B}_s^{*0} \to K^0\bar{D}^0)&=A_{K^0\bar{D}^0}V_{ub}V_{cd}^*\alpha_2\\
{\cal A}(\bar{B}_s^{*0} \to K^+D_s^-)&=A_{K^+D_s^-}V_{ub}V_{cs}^*\alpha_1\\
{\cal A}(\bar{B}_s^{*0} \to K^-D_s^+)&=A_{D_s^+K^-}V_{cb}V_{us}^*\alpha_1
\end{align}

\end{itemize}

Using the amplitudes given above, in the rest frame of $B^*$ meson, the branching fraction for $B^*\to M_1M_2$ decays can be written as 
\begin{equation}
{\cal B}(B^*\to M_1M_2)=\frac{1}{3}\frac{1}{8\pi}\frac{p_c}{m^2_{B^*}\Gamma_{tot}(B^*)}|{\cal A}(B^*\to M_1M_2)|^2\,,  \label{BR1}
\end{equation}
where, $\Gamma_{tot}(B^*)$ is the total decay width of $B^*$, and the momentum of final states is 
\begin{equation}
p_c=\frac{\sqrt{[m^2_{B^*}-(m_{M_1}+m_{M_2})^2][m^2_{B^*}-(m_{M_1}-m_{M_2})^2]}}{2m_{B^*}}\,.
\end{equation}

\section{Numerical Results and Discussion}
\subsection{Input papameters}
For the CKM matrix elements, we adopt the Wolfenstein parameterization~\cite{Wolfenstein:1983yz} and choose the four parameters $A$, $\lambda$, $\rho$ and $\eta$
as~\cite{Charles:2004jd}
\begin{equation}
A=0.810^{+0.018}_{-0.024}, \quad
\lambda=0.22548^{+0.00068}_{-0.00034}, \quad
\overline{\rho}=0.1453^{+0.0133}_{-0.0073}, \quad
\overline{\eta}=0.343^{+0.011}_{-0.012},
\end{equation}
with $\overline{\rho}=\rho\,(1-\frac{\lambda^2}{2})$ and
$\bar{\eta}=\eta\,(1-\frac{\lambda^2}{2})$.

As for the quark masses, we take~\cite{PDG14}
\begin{align}
 &\frac{\bar{m}_s(\mu)}{\bar{m}_q(\mu)} = 27.5 \pm 1, \quad
   \bar{m}_{s}(2\,{\rm GeV}) = 95 \pm 5 \,{\rm MeV}, \quad
   \bar{m}_{c}(\bar{m}_{c}) = 1.275 \pm 0.025 \,{\rm GeV},
   \nonumber \\
 &\bar{m}_{b}(\bar{m}_{b}) = 4.18 \pm 0.03 \,{\rm GeV}, \quad
  {m}_{t} = 173.21\pm0.51\pm0.71\,{\rm GeV},
\end{align}
 where $\bar{m}_q(\mu)=(\bar{m}_u+\bar{m}_d)(\mu)/2$,
and the difference between $u$ and $d$ quark is not distinguished. 

The decay constants of light mesons are \cite{PDG14}
\begin{align}
 f_{\pi} &= (130.41 \pm 0.02 \pm 0.20)\,{\rm MeV}, \quad
 &&f_{K} = (156.2\pm0.2\pm0.6 \pm 0.3)\,{\rm MeV}, \nonumber\\
 f_{D} &=(204.6\pm5.0)\,{\rm MeV}, \quad 
 &&f_{D_s} =(257.5\pm4.6)\,{\rm MeV},
\end{align}
which are extracted mainly from experimental measurements of some tree-dominated mesons decays~\cite{PDG14}. 

\begin{table}[htbp] \footnotesize
\caption{The predictions for radiative decay rates of $B_{u,d,s}^*$ mesons (in units of eV) within some theoretical approaches (LFQM: light front quark model; RQM: relativistic quark model; LCSR: light cone QCD sum rules; SR: QCD sum rules; HQET: Heavy Quark Effective Theory; VMD: Vector Meson Dominance hypothesis; CM: Covariant model).}
\label{tab:DW}
\centering   \vspace{0.5cm}
\begin{tabular}{c|ccccccc}
\hline\hline   
Decay modes  &   LFQM\cite{Choi:2007se}    &  RQM\cite{Goity:2000dk}   &  RQM\cite{Ebert:2002xz}  & LCSR\cite{Aliev:1995wi}  & SR\cite{Zhu:1996qy}  &   HQET+VMD\cite{Colangelo:1993zq}     &  HQET+CM\cite{Cheung:2014cka} \\
\hline
$B^{*+}\to B^+\gamma$  &   $400\pm 30$ &  $572^{+71}_{-65}$  &       190        &      1200       &    $380\pm60$ &      $220\pm90$     & $468^{+73}_{-75}$  \\%\vspace{1.0mm}
$B^{*0}\to B^0\gamma$  &   $130\pm 10$ &  $182^{+22}_{-21}$  &       70         &      280        &    $130\pm30$ &      $75\pm27$      &  $148\pm20$         \\%\vspace{1.0mm}
$B^{*0}_s\to B^0_s\gamma$  &   $68\pm 17$ &  $113\pm10$       &       54         &      --        &    $220\pm40$ &      --        &  --              \\
\hline \hline
\end{tabular}
\end{table}
To evaluate the branching fractions of $B^*_{(s)}$ weak decays, the total decay widths~(or lifetimes) $\Gamma_{tot}(B^*_{(s)})$ are essential. However, unfortunately,  there is no available experimental or theoretical information for $\Gamma_{tot}(B^*_{(s)})$ until now.  In our following numerical calculation, due to the known fact that the radiative  processes  $B^*_{(s)}\to B_{(s)}\gamma$ dominate the decays of $B^*_{(s)}$ mesons~\cite{PDG14}, the approximation $\Gamma_{tot}(B^*_{(s)})\simeq \Gamma(B^*_{(s)}\to B_{(s)}\gamma)$ are taken. The radiative decay rates of $B_{u,d,s}^*$ mesons have been widely studied in various theoretical models, whose theoretical predictions are summarized in Table~\ref{tab:DW}. Combining their ranges, we get
%From Table~\ref{tab:DW}, one may easily find that the results of LCSR\cite{Aliev:1995wi}~(RQM\cite{Ebert:2002xz}) are significantly larger~(smaller) than other ones. Averaging the residual results, we get 
\begin{align}
\label{eq:GamBp}
\Gamma(B^{*+}\to B^+\gamma)&=[130,640]\,{\rm eV}\,,\\
\label{eq:GamB0}
\Gamma(B^{*0}\to B^0\gamma)&=[50,280]\,{\rm eV}\,,\\
\label{eq:GamBs}
\Gamma(B^{*0}_s\to B^0_s\gamma)&=[50,260]\,{\rm eV}\,,
\end{align}
in which the result $\Gamma(B^{*+}\to B^+\gamma)=1200\,{\rm eV}$~(LCSR) is not considered because it is much larger than the other theoretical results. In the following numerical evaluations, such ranges are used and treated as $\Gamma_{tot}$.

Besides the input parameters given above, the transition form factor $A_0^{B^*_{(s)}\to M_1}(q^2)$ is also an essential ingredient for estimation of a certain nonleptonic $B^*_{(s)}$ decay. However, there is no available result until now, even though it could be calculated in many theoretical approaches. In this paper, we adopt Bauer-Stech-Wirbel (BSW) model~\cite{Wirbel:1985ji} to evaluate the values of form factors. At $q^2=0$,  $A_0^{B^*_{(s)}\to M_1}(0)$ is approximately equal to the overlap factor $h_{A_0}$, which could be written as~\cite{Wirbel:1985ji} 
\begin{align}
h_{A_0}=\int d^2p_{\perp}\int_{0}^{1} dx\, {\varphi_P^*(\vec{p}_\perp,x)\,\sigma_z^{(1)}\,\varphi^{1,0}_{V}(\vec{p}_\perp,x)}\,,
\end{align}
where $\vec{p}_\perp$ is the transverse quark momentum, $\sigma_z^{(1)}$ the Pauli matrix acting on the spin indices of the decaying quark. For the meson wave function $\varphi_M(\vec{p}_\perp,x)$, with the solution of a relativistic scalar harmonic oscillator potential, it is found to be~\cite{Wirbel:1985ji} 
\begin{equation}
\varphi_M(\vec{p}_\perp,x)=N_M\sqrt{x(1-x)}e^{-\vec{p}^2_\perp/2\omega^2}e^{-\frac{m^2}{2\omega^2}(x-\frac{m_M^2+m^2_1-m^2_2}{2m_M^2})^2}\,.
\end{equation}
In which, $N_M$ is the normalization factor; $m_{2(1)}$ represents the mass of (non-)spectator quark; and the parameter $\omega$ determines the average transverse quark momentum through $\langle p^2_\perp\rangle=\omega^2$.
%, which typical value is $\omega\simeq0.40~{\rm GeV}$ due to the well reproduction of experimental data~\cite{Wirbel:1985ji}
With the constituent masses $m_{u,d}=0.35~{\rm GeV}$, $m_{s}=0.55~{\rm GeV}$, $m_{c}=1.7~{\rm GeV}$, $m_{b}=4.9~{\rm GeV}$ and $\omega=0.40~{\rm GeV}$ as inputs, we get %the numerical results
\begin{align}\label{eq:ff}
&A_0^{B^*\to D}(0)=0.71\,, \quad A_0^{B_s^*\to D_s}(0)=0.66\,,\nonumber\\
&A_0^{B^*\to K}(0)=0.39\,, \quad A_0^{B^*\to \pi}(0)=0.34\,, \quad A_0^{B_s^*\to K}(0)=0.28\,.
\end{align}
It should be noted that the results of $A_0(0)$ are affected by some undetermined factors, such as the meson wave function and  the value of $\omega$~($\omega=0.4~{\rm GeV}$ is taken according to the data of D meson decays~\cite{Wirbel:1985ji}, but we do not know whether such value is suitable for $B^*$ decays). Conservatively, in our following evaluation, we take $15\%$ of the values in eq.~\eqref{eq:ff} as their uncertainties.

\begin{table}[] \footnotesize
\caption{ The numerical values of Wilson coefficients $C_i^{NLO}$ at $\mu=m_b$. }
\label{tab:WC}
\centering   \vspace{0.5cm}
\begin{tabular}{cccccccc}
\hline\hline
$C_1$ &   $C_2$    &  $C_4$   &  $C_6$  & $C_8/\alpha$  & $C_{10}/\alpha$  \\\hline
$1.084$ &   $-0.188$    &  $-0.036$   &  $-0.042$  & $0.056$  & $0.227$ \\
\hline \hline
\end{tabular}
\end{table}

In addition, for the other inputs, such as the masses of light mesons, the Fermi coupling constant $G_F$ and so on, we take their central values given in PDG~\cite{PDG14}.  Using the basic formulae given in Ref.~\cite{ref:Buras1,ref:Buras2} and the up-to-date inputs, we get the numerical value of Wilson coefficients $C_i$ relevant to our following evaluation and summarize them in Table~\ref{tab:WC}. 
\subsection{Numerical results}

 \begin{table}[t] \small
\caption{The branching fractions of $B^{\ast}\to M_1M_2$ decays.}
\centering  \vspace{0.5cm} \label{tab:Bd}
\begin{tabular}{ccc}
\hline\hline
    Decay Modes                                  &   Class      &   $\cal B$           \\
\hline
  $B^{*-} \to D^0D^-$                   &    T, P, ${\rm P_{ew}}$           & $[ 0.6, 3.7]\times 10^{-10}$   \\
  $\bar{B}^{*0} \to D^+D^-$             &    T, P, ${\rm P_{ew}}$     & $[ 1.5, 9.7]\times 10^{-10}$    \\
  $B^{*-} \to D^0D^-_s$                 &    T, P, ${\rm P_{ew}}$         & $[ 2, 10]\times 10^{-9}$   \\
  $\bar{B}^{*0} \to D^+D^-_s$           &    T, P, ${\rm P_{ew}}$    & $[ 0.4, 2.7]\times 10^{-8}$    \\
\hline
  $B^{*-} \to \pi^-D^0$                 &    T,C       & $[ 0.6, 3.9]\times 10^{-9}$  \\
  $B^{*-} \to \pi^-\bar{D}^0$           &    C      & $[ 0.5, 3.2]\times 10^{-14}$ \\
  $B^{*-} \to \pi^0D^-$                 &    T          & $[ 1.0, 5.7]\times 10^{-13}$    \\
  $\bar{B}^{*0} \to \pi^-D^+$           &    T     & $[ 2, 13]\times 10^{-9}$  \\
  $\bar{B}^{*0} \to \pi^+D^-$           &    T     & $[ 0.4, 3.0]\times 10^{-12}$  \\
  $\bar{B}^{*0} \to \pi^0D^0$           &    C     & $[ 2, 11]\times 10^{-11}$    \\
  $\bar{B}^{*0} \to \pi^0\bar{D}^0$     &    C     & $[ 0.7, 4.3]\times 10^{-14}$    \\
  $B^{*-} \to \pi^0D^-_s$               &    T     & $[ 3, 16]\times 10^{-12}$    \\
  $\bar{B}^{*0} \to \pi^+D^-_s$         &    T     & $[ 1.3, 8.4]\times 10^{-11}$ \\
\hline
  $B^{*-} \to K^-D^0$                   &    T,C   & $[ 0.5, 2.9]\times 10^{-10}$  \\
  $B^{*-} \to K^-\bar{D}^0$             &    C     & $[ 1.2, 7.8]\times 10^{-13}$  \\
  $\bar{B}^{*0}\to K^-D^+$              &    T     & $[ 1.5, 9.8]\times 10^{-10}$  \\
  $\bar{B}^{*0} \to \bar{K}^0D^0$       &    C     & $[ 2, 15]\times 10^{-12}$    \\
  $\bar{B}^{*0} \to \bar{K}^0\bar{D}^0$ &  C       & $[ 3, 20]\times 10^{-13}$    \\
\hline\hline
\end{tabular}
\end{table}

\begin{table}[ht] \small
\caption{The branching fractions of $\bar{B}^{\ast 0}_s\to M_1M_2$ decays.}
\centering  \vspace{0.5cm} \label{tab:Bs}
\begin{tabular}{ccc}
\hline\hline
   Decay Modes                               &   Class       &   $\cal B$            \\
\hline
   $\bar{B}^{*0}_s \to D^+_sD^-$     &    T, P, ${\rm P_{ew}}$     & $[ 1.4, 8.6]\times10^{-10}$   \\
   $\bar{B}^{*0}_s \to D^+_sD^-_s$   &    T, P, ${\rm P_{ew}}$     & $[ 0.4, 2.4]\times 10^{-8}$    \\
\hline
   $\bar{B}^{*0}_s \to \pi^-D^+_s$   &    T     & $[ 0.8, 4.6]\times 10^{-9}$   \\
\hline
   $\bar{B}^{*0}_s \to K^+D^-$       &    T     & $[ 3, 21]\times 10^{-13}$    \\
   $\bar{B}^{*0}_s \to K^0D^0$       &    C     & $[ 2, 15]\times 10^{-11}$  \\
   $\bar{B}^{*0}_s \to K^0\bar{D}^0$ &    C     & $[ 1.1, 6.0]\times 10^{-14}$  \\
   $\bar{B}^{*0}_s \to K^+D^-_s$     &    T     & $[ 0.9, 5.9]\times 10^{-11}$    \\
   $\bar{B}^{*0}_s \to K^-D^+_s$     &    T     & $[ 1.4, 8.7]\times 10^{-10}$    \\
\hline\hline
\end{tabular}
\end{table}

With the aforementioned values of input parameters and the theoretical formula in section 2, we then present our theoretical prediction and discussion. The ranges of branching fractions are given in Tables~\ref{tab:Bd} and \ref{tab:Bs}, in which the large uncertainties are mainly induced by $\Gamma(B^{*})$ and \eqref{eq:GamBs} and form factors. 
%In each table, the first, second and third theoretical errors are caused by uncertainties of the CKM parameters, hadronic parameters~(decay constants and form factors) and total decay width, respectively. Each of theoretical errors is  gotten by adding the uncertainties induced by different input parameters in quadrature.
The following are some discussions:  
%From Tables~\ref{tab:Bd} and \ref{tab:Bs}, it could be found that:
\begin{enumerate}
\item [(1)] There is a very clear hierarchy of the branching fractions of $B^*_{(s)}$ decays. For instance, for $B^*_{u,d}\to DD_{(s)}$ decays, ${\cal B}(\bar{B}^{*0} \to D^+D^-_s)$ $\sim$ $3\times{\cal B}(B^{*-} \to D^0D^-_s)$ $>$ ${\cal B}(\bar{B}^{*0} \to D^+D^-)$ $\sim$ $3\times{\cal B}(B^{*-} \to D^0D^-)$, which is caused by the following reasons: (i) The CKM factor $V_{cb}V_{cd}^*$ responsible for $B^{*-\,,0} \to D^{0\,,+}D^-$ decays is suppressed by a factor $\lambda$ relative to the CKM factor $V_{cb}V_{cs}^*$ responsible for $B^{*-\,,0} \to D^{0\,,+}D_s^-$ decays. In addition, $f_{D_s}/f_{D}\simeq1.258$~\cite{PDG14}. (ii) From Eqs.~(\ref{eq:bmd0dm}-\ref{eq:b0dpdsm}), one may find that ${\cal A}(\bar{B}^{*0} \to D^+D^-_{(s)})$ $\simeq$ ${\cal A}(B^{*-} \to D^0D^-_{(s)})$. Further, taking into account $\Gamma_{tot}(B^{*-})/\Gamma_{tot}(B^{*0})\sim 3$ illustrated by Eqs. \eqref{eq:GamBp} and \eqref{eq:GamB0}, ${\cal B}(\bar{B}^{*0} \to D^+D^-_{(s)})/{\cal B}(B^{*-} \to D^0D^-_{(s)})\sim 3$ could be easily understood. In fact, such  relationship between $B^*_{u,d}\to DD_{(s)}$ decays  could be expressed in a clearer way, 
\begin{align}
\frac{{\cal B}(\bar{B}^{*0} \to D^+D^-_{(s)})}{{\cal B}(B^{*-} \to D^0D^-_{(s)})}&\simeq\frac{\Gamma_{tot}(B^{*+})}{\Gamma_{tot}(B^{*0})}\,,\\
\frac{{\cal B}(\bar{B}^{*0} \to D^+D^-_{s})}{{\cal B}(\bar{B}^{*0} \to D^+D^-)}&\simeq \left|\frac{f_{D_s}}{f_{D}}\right|^2\left|\frac{V_{cs}}{V_{cd}}\right|^2\,.
\end{align}
The hierarchies or relationship in the other (color-suppressed) tree-dominated decay modes could be easily gotten through similar analysis. 

\item [(2)] In principle, $B^{*-} \to K^-\bar{D}^0$ decay is a theoretically clean channel to extract the CKM angle $\gamma$. Unfortunately, recalling our analysis of the capability of the experimental measurement at Belle-II in introduction, such color-suppressed decay mode is almost impossible to be measured due to its very small branching fraction $\sim {\cal O}(10^{-13})\ll{\cal O}(10^{-9})$. In fact, from the experimental point of view, most of decays calculated in this paper are hardly to be observed soon, except for a few tree-dominated decay modes with branching fractions $\gtrsim{\cal O}(10^{-9})$. So, in the following, only these possibly detectable decays are discussed. 

\item [(3)]  The CKM-favored and tree-dominated $\bar{B}^{*0} \to D^+D^-_s$ and $\bar{B}^{*0}_s \to D^+_sD^-_s$ decays have the largest branching fractions of the order $\sim{\cal O}(10^{-8})$ in $\bar{B}^{*0}$ and $\bar{B}^{*0}_s$ systems, respectively. Therefore, they should be sought for with priority and firstly observed at the running LHC and/or forthcoming Belle-II. 

\item [(4)] For such two easily detectable decay modes, the $SU(3)$ symmetry (or U-spin symmetry acting on the spectator of $B^{*}$ meson) implies the relation
\begin{align}
{\cal A}(\bar{B}^{*0} \to D^+D^-_s)\simeq {\cal A}(\bar{B}^{*0}_s \to D^+_sD^-_s)\,,
\end{align}
which is satisfied in our numerical evaluation. With that, the ratio defined as
\begin{align}\label{eq:rd}
R_D \equiv\frac{{\cal B}(\bar{B}^{*0} \to D^+D^-_s)}{{\cal B}(\bar{B}^{*0}_s \to D^+_sD^-_s)}
\end{align}
would be a useful observable to measure $\tau_{\bar{B}^{*0}_d}/\tau_{\bar{B}^{*0}_s}$~($\tau_{\bar{B}^{*0}_{d,s}}$ are the lifetimes of $\bar{B}^{*0}_{d,s}$ mesons), which $\simeq R_D$. Moreover, if $\Gamma_{tot}(B^*_{(s)})\simeq \Gamma(B^*_{(s)}\to B_{(s)}\gamma)$ is a good approximation, it could be further used to test the results of $\Gamma(B^*_{(s)}\to B_{(s)}\gamma)$ calculated in various models listed in Table~\ref{tab:DW}. For instance, $\Gamma(B^{*0}_s\to B^0_s\gamma)/\Gamma(B^{*0}\to B^0\gamma)\simeq 1.7>1$ in SR~\cite{Zhu:1996qy} is obviously different from the results $\sim [0.5,0.8]<1$ gotten in the other models~\cite{Choi:2007se,Goity:2000dk,Ebert:2002xz}~(see Table~\ref{tab:DW} for detail).
A similar case also exists in  $\bar{B}^{*0} \to \pi^-D^+$  and $\bar{B}^{*0}_s \to \pi^-D^+_s$ decay modes.

\item [(5)]  For $B^{*-} \to \pi^-D^0$ decay, it should be noted that there is a significant cancellation effect between the tree term related to $\alpha_1$ and the color-suppressed one related to $\alpha_2$ in Eq.~\eqref{BmpimD0}, which is different from the situation in $B\to PV(VP)$ decays. It is caused purely by a kinematic reason that $\varepsilon \cdot p_{\pi^-}=-\varepsilon \cdot p_{D^0}=p_c$~(or $-p_c$)~\footnote{In Eq.~\eqref{BmpimD0}, $\varepsilon \cdot p_{\pi^-}$ involved in $A_{D^0\pi^-}$ and $\varepsilon \cdot p_{D^0}$ involved in $A_{\pi^-D^0}$ could be replaced by $p_c$ and $-p_c$~(or $-p_c$ and $p_c$), respectively, since the amplitude squared and summed over the polarization of $B^*$ meson gives $|p_c|^2$ for $|A_{D^0\pi^-}|^2$ and $|A_{\pi^-D^0}|^2$ terms, but $-|p_c|^2$ for interference terms.}. As a result, ${\cal B}(B^{*-} \to \pi^-D^0)/{\cal B}(\bar{B}^{*0} \to \pi^-D^+)<\tau_{B^{*-}}/\tau_{\bar{B}^{*0}}$, which is equal to $\tau_{B^{*-}}/\tau_{\bar{B}^{*0}}$ in absence of color-suppressed contribution, is required.

\item[(6)]  The predictions in Tables~\ref{tab:Bd} and \ref{tab:Bs} are based on the NF approximation, which could be treated as the leading order~(LO) results of QCDF. The QCD contributions at next-to-leading order~(NLO) and even next-to-next-to-leading order~(NNLO), which is essential for a more reliable results  and has attract much attention~\cite{Beneke1,Beneke:2003zv,Beneke:2001ev,Beneke:2005vv,Beneke:2006mk,Beneke:2009ek}, may provide sizable corrections to the LO.  For example, in the tree-dominated $B\to\pi\pi$ decays, the color-allowed and -suppressed  tree amplitudes at NLO read~\cite{Beneke1,Beneke:2009ek}
\begin{align}
\alpha_{1}(\pi\pi)&= [1.009]_{LO}+[0.023+0.010i]_{NLO}^V-\left(\frac{r_{sp}}{0.445}\right)[0.014]_{LOsp}\,,\nonumber\\
\alpha_{2}(\pi\pi)&= [0.220]_{LO}-[0.179+0.077i]_{NLO}^V+\left(\frac{r_{sp}}{0.445}\right)[0.114]_{LOsp}\,,\nonumber
\end{align}
respectively. It is found that: (i) For the color-allowed tree amplitude $\alpha_{1}$,  relative to the LO amplitude, the ${\cal O}(\alpha_s)$ corrections are numerically trivial.  (ii) For the color-suppressed tree amplitude $\alpha_{2}$, the one-loop vertex correction is sizable, about $89\%$ compared with the LO amplitude.  While, when the spectator-scattering correction is included, the ${\cal O}(\alpha_s)$ correction is reduced to about $46\%$ due to the cancellation. In addition, the overall NNLO correction, already known from Refs.~\cite{Beneke:2005vv,Beneke:2009ek}, to the topological tree amplitudes is small~\cite{Beneke:2009ek}. Assuming such findings hold roughly in the tree-dominated $B^*$ decays, the NF estimates for the color-allowed  tree-dominated decay modes are relatively stable due to the small QCD corrections. For the color-suppressed tree-dominated $B^*$ decays, even though the NF results possibly suffer significant correction from QCD, these decay modes still escape the experimental scope because of their very small branching fraction, $\lesssim{\cal O}(10^{-11})$. 
\end{enumerate}

\section{Summary}
Motivated by the future heavy flavor physics experiments at running LHC and upgrading SuperKEKB/Belle-II with high-luminosity, the nonleptonic $B^{\ast}_{(s)}\to M_1 M_2$ $(M=D$, $D_s$, $\pi$, $K)$ weak decays are studied in detail. After a brief review of the effective Hamiltonian and factorization approach, some generic formula and amplitudes of  $B^{\ast}_{(s)}\to M_1 M_2$ decays are presented. With the reasonable approximation $\Gamma_{tot}(B^*_{(s)})\simeq \Gamma(B^*_{(s)}\to B_{(s)}\gamma)$, we have further presented their numerical results of branching fractions in Tables~\ref{tab:Bd} and \ref{tab:Bs}, in which the transition form factors $A_0^{B^{\ast}_{(s)}\to M_1}(0)$ are calculated within the BSW model. It is found that most of the decay modes evaluated in this paper are too rare to be observed soon, except for some tree-dominated and CKM-favored ones with branching fractions~$\gtrsim{\cal O}(10^{-9})$. In which,  $\bar{B}^{*0} \to D^+D^-_s$ and $\bar{B}^{*0}_s \to D^+_sD^-_s$ decays have the largest branching fractions of the order $\sim{\cal O}(10^{-8})$, and hence should be sought for with priority and firstly observed at the running LHC and/or forthcoming SuperKEKB/Belle-II. Besides, for the possible detectable $B^{\ast}_{(s)}$ decays, which branching fractions $\gtrsim{\cal O}(10^{-9})$, some useful ratios, such as $R_D$ defined by Eq.~\eqref{eq:rd}, are presented and discussed in detail.

\section*{Acknowledgments}
 The work is supported by the National Natural Science Foundation of China (Grant Nos. 11475055 and 11105043) and the Foundation for the Author of National Excellent Doctoral Dissertation of P. R. China (Grant No. 201317) and the Program for Science and Technology Innovation Talents in Universities of Henan Province (Grant No. 14HASTIT036). Q. Chang is also supported by the Funding Scheme for Young Backbone Teachers of Universities in Henan Province (Grant No. 2013GGJS-058). We also thank the Referees for their helpful comments.


\begin{thebibliography}{99}
 \bibitem{ref:LHCb2013}
 % arXiv:1208.3355
R. Aaij {\it et al.} (LHCb Collaboration), Eur. Phys. J. C {\bf 73} (2013) 4, 2373.

 \bibitem{ref:BelleII2011}
T. Abe {\it et al.} (Belle II Collaboration), arXiv:1011.0352.
 
\bibitem{PDG14}
  K. A. Olive {\it et al.}  (Particle Data Group),  Chin. Phys. C {\bf 38} (2014) 090001.

%\cite{Huang:2006em}
\bibitem{Huang:2006em}
  %``Measurement of B(Upsilon(5S) ---> B(*)(s) anti-B(*)(s)) Using phi Mesons,''  [hep-ex/0610035]
  G.~S.~Huang {\it et al.}  (CLEO Collaboration),  Phys.\ Rev.\ D {\bf 75} (2007) 012002.

\bibitem{Louvot:2008sc}
  %``Measurement of the Decay $B_s 0 \to D_s - \pi^{+}$ and Evidence for $B_s 0 \to D_s \pm K^\pm$ in $e^+ e_-$ Annihilation at $\sqrt{s}$ ~ 10.87-GeV,'' [arXiv:0809.2526 [hep-ex]]
  R.~Louvot {\it et al.}  (Belle Collaboration),  Phys.\ Rev.\ Lett.\  {\bf 102} (2009) 021801.

\bibitem{Louvot:2009ie}
  R.~Louvot (Belle Collaboration),   PoS EPS {\bf -HEP2009} (2009) 170.
  %``B0(S) Decays at Belle,'' [arXiv:0909.2160 [hep-ex]].

\bibitem{Aquines:2006qg}
  %``First measurements of the exclusive decays of the upsilon(5S) to B meson final states and improved B*(S) mass measurement,'' [hep-ex/0601044].
  O.~Aquines {\it et al.}  (CLEO Collaboration),  Phys.\ Rev.\ Lett.\  {\bf 96} (2006) 152001.

\bibitem{Aaij:2012uva}
  R.~Aaij {\it et al.}  (LHCb Collaboration),  Phys.\ Rev.\ Lett.\  {\bf 110} (2013) 15,  151803.
  %``First observation of the decay $B_{s2}^*(5840)^0 \to B^{*+} K^-$ and studies of excited $B^0_s$ mesons,''[arXiv:1211.5994 [hep-ex]]

\bibitem{Abe:2010gxa}
  %``Belle II Technical Design Report,''
  T.~Abe {\it et al.}  (Belle-II Collaboration), arXiv:1011.0352.

\bibitem{Huang:2006mf}
  %``Measurement of B(Upsilon(5S) ---> B(*)(s) anti-B(*)(s)) Using phi Mesons,'' hep-ex/0607080.
  G.~S.~Huang {\it et al.}  (CLEO Collaboration), hep-ex/0607080.

\bibitem{Aaij:2010gn}
  %``Measurement of $\sigma(pp \to b \bar{b} X)$ at $\sqrt{s}=7~\rm{TeV}$ in the forward region,''  [arXiv:1009.2731 [hep-ex]]
  R.~Aaij {\it et al.}  (LHCb Collaboration),  Phys.\ Lett.\ B {\bf 694} (2010) 209.
   
\bibitem{Aaij:2014jba}
  %``LHCb Detector Performance,''[arXiv:1412.6352 [hep-ex]].
  R.~Aaij {\it et al.}  (LHCb Collaboration), Int.\ J.\ Mod.\ Phys.\ A {\bf 30} (2015) 07,  1530022.
    
\bibitem{Wang:2012hu}
   %``Semileptonic decays $B_c^* \to \eta_c \ell \bar{\nu}_{\ell} $ with QCD sum rules,''  [arXiv:1209.1157 [hep-ph]]
 Z.~G.~Wang,  Commun.\ Theor.\ Phys.\  {\bf 61} (2014) 1,  81.
   
\bibitem{Bashiry:2014qia}
%``Investigation of the rare exculsive $B_c^\ast \rightarrow D_{s}~\nu \bar{\nu}$ decays in the faramework of the QCD sum rules,'' [arXiv:1410.0529 [hep-ph]]
   V.~Bashiry, Adv.\ High Energy Phys.\  {\bf 2014} (2014) 503049.

\bibitem{Zeynali:2014wya}
  %``Form factors and decay rate of B$_{c}^{*}$ $ \rightarrow$ D$_{s}$l$^{+}$l$^{-}$ decays in the QCD sum rules,''  [arXiv:1410.0526 [hep-ph]].
  K.~Zeynali, V.~Bashiry and F.~Zolfagharpour,  Eur.\ Phys.\ J.\ A {\bf 50} (2014) 127.

\bibitem{ref:Buras1}
 G.~Buchalla, A.~J.~Buras, and M.~E.~Lautenbacher, Rev.\ Mod.\ Phys.\ {\bf 68} (1996) 1125. % [hep-ph/9512380];
 
 \bibitem{ref:Buras2}
 A.~J.~Buras, hep-ph/9806471.

\bibitem{Fakirov:1977ta}
  %``F and D Decays,''
  D.~Fakirov and B.~Stech,  Nucl.\ Phys.\ B {\bf 133} (1978) 315.
  
\bibitem{Bauer:1984zv}
  %``Exclusive d Decays,''
  M.~Bauer and B.~Stech,  Phys.\ Lett.\ B {\bf 152} (1985) 380.
  
\bibitem{Wirbel:1985ji}
  %``Exclusive Semileptonic Decays of Heavy Mesons,''
  M.~Wirbel, B.~Stech and M.~Bauer, Z.\ Phys.\ C {\bf 29} (1985) 637.
          
\bibitem{Bauer:1986bm}
  %``Exclusive Nonleptonic Decays of D, D(s), and B Mesons,''
  M.~Bauer, B.~Stech and M.~Wirbel,  Z.\ Phys.\ C {\bf 34} (1987) 103.
 
 \bibitem{Bjorken:1988kk}
  %``Topics in B Physics,''
  J.~D.~Bjorken,  Nucl.\ Phys.\ Proc.\ Suppl.\  {\bf 11} (1989) 325.
  
  \bibitem{Jain:1995dd}
 %``Quantum color transparency and nuclear filtering,''   [hep-ph/9511333]
  P.~Jain, B.~Pire and J.~P.~Ralston, Phys.\ Rept.\  {\bf 271} (1996) 67.
  
\bibitem{Beneke:2000wa}
  M.~Beneke and T.~Feldmann,
  %``Symmetry breaking corrections to heavy to light B meson form-factors at large recoil,''
  Nucl.\ Phys.\ B {\bf 592} (2001) 3.
  %[hep-ph/0008255].

 \bibitem{Beneke1}
  M. Beneke, G. Buchalla, M. Neubert  and C. Sachrajda,  Phys. Rev. Lett. {\bf 83} (1999) 1914. % [hep-ph/9905312];
  
   \bibitem{Beneke2}
  M. Beneke, G. Buchalla, M. Neubert  and C. Sachrajda, Nucl. Phys. B {\bf 591} (2000) 313. % [hep-ph/0006124].

 \bibitem{KLS1}
  Y. Keum, H. Li and A. Sanda,
  Phys. Lett. B {\bf 504} (2001) 6. % [hep-ph/0004004];
  
   \bibitem{KLS2}
    Y. Keum, H. Li and A. Sanda,
    Phys. Rev. D {\bf 63} (2001) 054008. % [hep-ph/0004173].

  \bibitem{scet1}
  C. Bauer, S. Fleming and M. Luke,
  Phys. Rev. D {\bf 63} (2000) 014006.% [hep-ph/0005275];
  \bibitem{scet2}
  C. Bauer, S. Fleming, D. Pirjol and I. Stewart,
  Phys. Rev. D {\bf 63} (2001) 114020.% [hep-ph/0011336];
  \bibitem{scet3}
  C. Bauer and I. Stewart,
  Phys. Lett. B {\bf 516} (2001) 134.% [hep-ph/0107001 ];
  \bibitem{scet4}
  C. Bauer, D. Pirjol and I. Stewart,
  Phys. Rev. D {\bf 65} (2002) 054022.% [hep-ph/0109045].

\bibitem{Beneke:2003zv}
  %``QCD factorization for B ---> PP and B ---> PV decays,'' [hep-ph/0308039]
M.~Beneke and M.~Neubert,  Nucl.\ Phys.\ B {\bf 675} (2003) 333. 

 \bibitem{Wolfenstein:1983yz}
 L.~Wolfenstein, Phys.\ Rev.\ Lett.\ {\bf 51} (1983) 1945.

\bibitem{Charles:2004jd}
 %[hep-ph/0406184]
 J.~Charles {\it et al.} (CKMfitter Group), Eur.\ Phys.\ J.\ C {\bf 41} (2005) 1;
 updated results and plots available at: http://ckmfitter.in2p3.fr..
 
  
\bibitem{Choi:2007se}
  %``Decay constants and radiative decays of heavy mesons in light-front quark model,'' [hep-ph/0701263]
  H.~M.~Choi, Phys.\ Rev.\ D {\bf 75} (2007) 073016.
  
\bibitem{Goity:2000dk}
  %``Radiative transitions in heavy mesons in a relativistic quark model,''  [hep-ph/0012314]
  J.~L.~Goity and W.~Roberts, Phys.\ Rev.\ D {\bf 64} (2001) 094007.
  
\bibitem{Ebert:2002xz}
  %``Radiative M1 decays of heavy light mesons in the relativistic quark model,''  [hep-ph/0204089]
  D.~Ebert, R.~N.~Faustov and V.~O.~Galkin,  Phys.\ Lett.\ B {\bf 537} (2002) 241.
  
\bibitem{Aliev:1995wi}
  %``Radiative B* ---> B gamma and D* gamma decays in light cone QCD sum rules,''  [hep-ph/9511362, hep-ph/9511290]
  T.~M.~Aliev, D.~A.~Demir, E.~Iltan and N.~K.~Pak,  Phys.\ Rev.\ D {\bf 54} (1996) 857.
  
\bibitem{Zhu:1996qy}
  %``D* ---> D gamma and B* ---> B gamma as derived from QCD sum rules,''  [hep-ph/9610412]
  S.~L.~Zhu, W.~Y.~P.~Hwang and Z.~S.~Yang,  Mod.\ Phys.\ Lett.\ A {\bf 12} (1997) 3027.
  
\bibitem{Colangelo:1993zq}
  %``Radiative heavy meson transitions,'' [hep-ph/9307330]
  P.~Colangelo, F.~De Fazio and G.~Nardulli,  Phys.\ Lett.\ B {\bf 316} (1993) 555.
  
\bibitem{Cheung:2014cka}
  %``Strong and radiative decays of heavy mesons in a covariant model,''  [arXiv:1401.3917 [hep-ph]]
  C.~Y.~Cheung and C.~W.~Hwang,  JHEP {\bf 1404} (2014) 177.
  
\bibitem{Beneke:2001ev}
  %``QCD factorization in B ---> pi K, pi pi decays and extraction of Wolfenstein parameters,'' [hep-ph/0104110].
M.~Beneke, G.~Buchalla, M.~Neubert and C.~T.~Sachrajda, Nucl.\ Phys.\ B {\bf 606} (2001) 245. 
 

\bibitem{Beneke:2005vv}
 %``Spectator scattering at NLO in non-leptonic b decays: Tree amplitudes,''  [hep-ph/0512351].
   M.~Beneke and S.~Jager, Nucl.\ Phys.\ B {\bf 751} (2006) 160.

\bibitem{Beneke:2006mk}
  %``Spectator scattering at NLO in non-leptonic B decays: Leading penguin amplitudes,'' [hep-ph/0610322].
  M.~Beneke and S.~Jager, Nucl.\ Phys.\ B {\bf 768} (2007) 51. 
 
\bibitem{Beneke:2009ek}
  %``NNLO vertex corrections to non-leptonic B decays: Tree amplitudes,''  [arXiv:0911.3655 [hep-ph]].
 M.~Beneke, T.~Huber and X.~Q.~Li,  Nucl.\ Phys.\ B {\bf 832} (2010) 109.  

\end{thebibliography}
 \end{document}